\documentstyle [aps] {revtex}
\begin {document}
\title {Black hole entropy from Poisson brackets \\ (demystification of some calculations)}
 \author {Vladimir O. Soloviev \\
 {\small \it Institute for High Energy Physics} \\
 {\small \it 142 284, Protvino, Moscow region, Russia}\\
{\small \it e-mail: soloviev@th1.ihep.su}}
\maketitle

\begin {abstract}
Recently it has been suggested by S. Carlip that black hole
entropy can be derived from a central charge of the Virasoro
algebra arising as a subalgebra in the surface deformations of
General Relativity in any dimension. Here it is shown that the
argumentation  based on
the Regge-Teitelboim approach is unsatisfactory. The functionals
used are really ``non-differentiable'' under required variations
and also the standard Poisson brackets for these functionals are
exactly zero so being unable to get any Virasoro algebra with a
central charge. Nevertheless Carlip's calculations will be correct
if we admit another definition for the Poisson bracket. This new
Poisson bracket differs from the standard one in surface terms
only and allows to work with  ``non-differentiable'' functionals.
\end {abstract}

\section {Introduction}
An important for black hole physics result --- the entropy formula --- was derived in publication \cite{Carlip}. The similar conclusions were obtained also in some other papers, see, for example, \cite{Solod}. The characteristic feature of the method applied in \cite{Carlip} is that  the canonical formalism of General Relativity and the algebra of the hypersurface deformations are used. The Virasoro  algebra with its central term arises as a result of the evaluation of Poisson brackets and taking into account boundary conditions on the black hole horizon.

In \cite{Carlip} it is declared  that the Regge-Teitelboim approach based on a construction of the so-called ``differentiable'' generators is used there. However, a detailed examination  shows that this approach is really inaplicable to the case under consideration. The generators of paper \cite{Carlip} actually are not ``differentiable'' concerning variations needed to derive the Virasoro albebra.  In fact, the straightforward and explicit calculation of the Poisson brackets according to the standard formula (by exploiting the general derivation of the surface terms made for 3+1--space-time still in 1985) gives us a trivial zero result for the surface integral which should represent the main outcome of paper \cite{Carlip}.

We will demonstrate below that in fact  modified Poisson brackets, applicable in the more general case of ``non-differentiable'' functionals, are used in publication \cite{Carlip} (but without author's knowledge). These new brackets differ from the standard ones in boundary terms and they have been used for the first time in paper \cite{LMMR}. Later they were generalized in two different ways \cite{Sol93}, \cite{Bering}. The difference between these two approaches does not manifest itself in the case under consideration. Recently the same brackets were independently used also in publications \cite{Park}, \cite{Hwang}.

Therefore we observe that the derivation of the black hole entropy
given in paper \cite{Carlip} formally stays valid if one changes
the argumentation. It is necessary to decline the Regge-Teitelboim
ideology here in favor of the more general approach which exploits
the new Poisson brackets.

\section {Notations and Carlip's Calculation} We shall use the same notations
as used in publication \cite {Carlip}. So, the metric of $n $-dimensional (in further  $n=4$ case is preferred) space-time will be guessed looking as follows
\[
ds^2 = -N^2 dt^2 + f^2 (dr + N^r dt)^2
+ \sigma_{\alpha\beta} (dx^\alpha + N^\alpha dt)(dx^\beta + N^\beta dt),
\]
where $ \alpha, \beta\ldots $ correspond to coordinates on a sphere
$r={\rm const},  t={\rm const}$ and run over values $ 1,2, \ldots, n-2 $. The function $N $ tends to zero on the  horizon of a black hole $r=r _ + $ in such a way that
\begin{equation}
N^2 = h (x^\alpha) (r-r_+) + O(r-r_+)^2.
\end{equation}

Further, the Hamiltonian looks like a linear combination of constraints
$ \{ {\cal H}_t, {\cal H}_a \} $  plus, probably, some surface integrals
\begin{equation}
H [{\hat\xi}] = \int_\Sigma d ^ {n-1} x \, {\hat\xi} ^ \mu {\cal H} _ \mu,\qquad
L[\hat\xi]=H[\hat\xi] + \oint _ {\partial\Sigma} \ldots
,
\end {equation}
where the parameters of deformations of a constant time hypersurface
(Lagrange multipliers for constraints) are  components of the decomposition for
an infinitesimal space-time diffeomorphism  $ \xi^\mu $ over the  basis $\{ (1/N, -N^a/N), \partial/\partial x^a \} $:
\begin{equation}
{\hat\xi}^t = N\xi^t, \qquad
{\hat\xi} ^a = \xi^a + N^a\xi^t.
\end {equation}
If one calculate a variation over $g _ {ab} $ and $ \pi ^ {ab} $ of that part of
a Hamiltonian containing only constraints (without the boundary terms),
and  consider $ \xi ^\mu $ as not depending of canonical variables, one get
\begin{eqnarray}
\delta H&=&\delta\int_\Sigma d^{n-1}x  {\hat\xi}^\mu {\cal H}_\mu =\nonumber\\
&=& \int_\Sigma d^{n-1}x\left(\frac{\delta H}{\delta g_{ab}} \delta g_{ab} + \frac{\delta H}{\delta\pi^{ab}} \delta\pi^{ab} \right) -\nonumber\\
&-&{1\over 16\pi G} \oint_{\partial\Sigma} d^{n-2}x  \left\{ \sqrt{\sigma} \left(\sigma^{ac} n^b -\sigma^{ab} n^c\right)\left({\hat\xi}^t\nabla_c\delta g_{ab}- \nabla_c {\hat\xi}^t\delta g_{ab} \right)+ 2 {\hat\xi}^a\delta\pi_a^{\ r} - {\hat\xi}^r \pi^{ab} \delta g_{ab}\right\},\label{eq:variation}
\end{eqnarray}
where $n^a $ is  a unit normal to the boundary at $t = {\rm const} $, $K _{ab} $ is an extrinsic curvature tensor  of a constant time hypersurface, $\pi^{ab} = - f\sqrt {\sigma} (K^{ab} - g^{ab} K) $ are momenta conjugate to the spatial metric $g _ {ab} $  (the sign differs from \cite {Carlip}). The surface integral is taken over a boundary which should include, on the one hand, the horizon of a black hole, and on another -- the spatial infinity. No boundary conditions are applied here. To within notations the formula coincides with the similar formula from publication by Regge and Teitelboim \cite {RT}.

In further the game is entered by boundary conditions on the canonical variables and on the parameters of deformations. They are
\begin {enumerate}
\item to set in a phase space a region ``close''  to a black hole solution;
\item to provide conservation of this region under evolution assigned to the
parameters of deformations.
\end {enumerate}
Carlip starts with the following boundary conditions:
\[
f = {\beta h\over4\pi} N ^ {-1} + O (1), \qquad
N^r = O (N^2), \qquad
\sigma _{\alpha\beta} = O (1), \qquad
N ^\alpha = O (1), \]
\[
(\partial_t -N^r\partial_r) g _ {\mu\nu} = O (N) g _ {\mu\nu}, \qquad
\nabla_\alpha N_\beta
+ \nabla_\beta N_\alpha = O (N), \qquad
\partial_r N = O (1/N), 
\] 
\[
K_ {rr} = O (1/N^3), \qquad
K _ {\alpha r} = O (1/N), \qquad
K _ {\alpha\beta} =O (1), 
\]
\[
{\hat\xi} ^r = O (N^2), \qquad
{\hat\xi} ^t = O (N), \qquad
{\hat\xi} ^ \alpha= O (1). 
\]
But a bit later (after eq.~(2.12)) the extra restrictions $K _ {rr} = 0 =K _ {\alpha\beta} $ are imposed on the exterior curvature tensor  in \cite {Carlip}. As a result the last term in the variation formula (\ref{eq:variation}) becomes exactly zero  and so we, in fact, need not add the term written in square brackets of (\ref{eq:correction}) below.  
What about a part of the boundary arranged at spatial infinity, by considering  parameters $\hat\xi $ as rapidly decreasing we get rid of the corresponding contribution to the Hamiltonian  variation.

Despite of the accepted boundary conditions the variation of Hamiltonian $H[\hat\xi]$ still contains some surface contribution and to get rid of it in publication \cite {Carlip} it is suggested to add to a linear combination of constraints a surface integral of the following form
\begin {equation}
J [{\hat\xi}] = {1\over8\pi G} \oint _ {r=r _ +}  d ^ {n-2} x \Bigl\{n^a\nabla_a {\hat\xi} ^t\sqrt {\sigma} + {\hat\xi}
^a\pi_a {} ^r + \left[n_a {\hat\xi} ^a K\sqrt {\sigma} \right]
\Bigr\}. \label{eq:correction}
\end {equation}
Let us immidiately state that the  term in square brackets is absolutely not required by the Regge-Teitelboim ideology, just opposite, it spoils the ``differentiability'' condition (in \cite{Carlip} this is masked by an unjustified restriction $\delta K_{rr}/K_{rr}=O(N)$), so we treat it as erroneous and it will be excluded from now on.
Then the variation of the improved Hamiltonian $L[\hat\xi]=H [{\hat\xi}] + J [{\hat\xi}]$ becomes
\begin {equation}
\delta L[\hat\xi] = \hbox {\em bulk terms}
+ {1\over8\pi G} \oint _ {r=r _ +}  d ^ {n-2} x \ \ \delta
n^r\partial_r {\hat\xi} ^t  \sqrt{\sigma}.
\end {equation}

Afterwards in publication \cite {Carlip} it is stated that,  first, we can
calculate a Poisson bracket for generators of two various deformations according
to the following formula \begin {equation}
\left\{L [{\hat\xi} _2], L [{\hat\xi} _1] \right\} = \delta _ {{\hat\xi} _2}
L [{\hat\xi} _1],\label{eq:7}
\end {equation}
and second, it is supposed, that this equality remain valid after reduction,
that is, after taking constraints as zeros ``in strong sense''. The last means, that the formula remains valid for surface integrals taken separately:
\begin {equation}
\hbox {\em boundary terms of} \left\{L [{\hat\xi} _2], L [{\hat\xi} _1]
\right\} \approx  \hbox {\em boundary terms of} \ \ \delta _ {{\hat\xi} _2}H[\hat\xi_1]+\delta _ {{\hat\xi} _2}J[{\hat\xi} _1]. \label{eq:8}
\end {equation}
As a result of such evaluation the following
expression was obtained for the surface contribution to the Poisson
bracket $ \left\{L [{\hat\xi} _2], L [{\hat\xi} _1] \right\} $:
\begin {equation}
- {1\over8\pi G} \oint _ {r=r _ +}  d ^ {n-2} x \, \sqrt {\sigma} \left\{
{ 1\over f^2} \left[\partial_r(f{\hat\xi}_2^r)\partial_r{\hat\xi}_1^t-
\partial_r(f{\hat\xi}_1^r)\partial_r{\hat\xi}_2^t \right]
+ {1\over f} \partial_r\left[{\hat\xi} _1^r\partial_r {\hat\xi} _2^t
- \delta _ {{\hat\xi} _2} {\hat\xi} _1^t \right]
\right\}.
\label {eq:2.13}
\end {equation}
In further it is used  to realize the Virasoro algebra, which central charge
has allowed to connect the number of horizon  states with the
well-known expression for the black hole entropy by Bekenstein-Hawking. Below we shall show,
that the argumentation of paper \cite {Carlip} on a derivation of this formula
should be reconsidered.

\section {The Regge-Teitelboim Approach} In publication by  Regge and
 Teitelboim \cite {RT}, written in 1974, the Hamiltonian formalism of a field
 theory for the first time was applied to a problem, where the surface integrals
 originating in integration by parts were non-negligible, just opposite, they have
 an important physical meaning. Namely, after putting on gauges, that is, after
 reduction, the role of generators of the asymptotic Poincar\'e  group
 (transformations that preserve boundary conditions of the  asymptotically flat
 space - time) is played  by the surface integrals. These surface integrals
 arise as a result of the Poisson brackets evaluation made according to the well-known
formula
\[
 \{F+\oint_{\partial\Sigma}\! \ldots,G+\oint_{\partial\Sigma}\! \ldots\}_{\rm old} =
\int_\Sigma d ^ {n-1} x\left(\frac {\delta F} {\delta q_A (x)} \frac {\delta
G} {\delta p_A (x)} - \frac {\delta F} {\delta p_A (x)} \frac {\delta G}
{\delta q_A (x)} \right),
\]
and any modification of functionals $F $ and $G $ by surface integrals $ \oint
_ {\partial\Sigma} \ldots $ (or, that is the same, a modification of their
 integrands by  divergences) leaves this bracket untouched. It is obvious, as
 the standard bracket is defined with the help of  Euler-Lagrange  derivatives, and they are zero at any divergences. It is a common belief that the essence of Regge-Teitelboim's method is to fix surface terms in
 functionals in such a way that their variations look like
\[
\delta F =\int_\Sigma d ^ {n-1} x\frac {\delta F} {\delta\phi_A (x)}
\delta\phi_A (x),
\]
without any boundary terms.

 However, in the Regge-Teitelboim  approach we can start also with the evaluation of surface integrals in Poisson brackets.
 This possibility was used in our publication \cite {Sol85}. Thus taking as initial Hamiltonians
\begin{equation}
H[{\hat\xi}] = \int_\Sigma d^{n-1}x\, {\hat\xi}^\mu{\cal H}_\mu ,
\end{equation}
where $\{ {\cal H}_t, {\cal H}_a \}$ are constraints, we obtain (we write below the corresponding formula from publication \cite{Sol85} in  the notations of Carlip's work (here $n=4$)) the following expression for their Poisson bracket:
\newpage 
\begin{eqnarray}
\{H[\hat\xi_1]+\oint_{\partial\Sigma} \ldots,H[\hat\xi_2]+\oint_{\partial\Sigma}
\ldots\}_{\rm old}&=&H[\hat\xi_3]+\nonumber\\
&+&\frac{1}{8\pi G}\oint_{\partial\Sigma} \hat\xi_3^a\pi_a^{\ b}dS_b+\oint_{\partial\Sigma}{\cal
H}_t(\hat\xi_1^t\hat\xi_2^a-\hat\xi_2^t
\hat\xi_1^a)dS_a-\nonumber\\
&-&\frac{1}{16\pi G}\oint_{\partial\Sigma}
\pi^{ab}\left[\hat\xi_1^c(\hat\xi_{2a|b}+
\hat\xi_{2b|a})-\hat\xi_2^c(\hat\xi_{1a|b}+\hat\xi_{1b|a})\right]dS_c+\nonumber\\
&+&\frac{1}{8\pi G}\oint_{\partial\Sigma}\sqrt{g}R^a_b\left(\hat\xi_1^t\hat\xi_2^b-\hat\xi_2^t\hat\xi_1^b\right)dS_a+\nonumber\\
&+&\frac{1}{8\pi G}\oint_{\partial\Sigma}\sqrt{g}(g^{ab}
g^{cd}-g^{ac}g^{bd})\left(\hat\xi_{1a}
\hat\xi^t_{2|bd}-\hat\xi_{2a}\hat\xi^t_{1|bd}\right)dS_c,
\end{eqnarray}
and
\begin{equation}
\hat\xi_3=\{\hat\xi_1,\hat\xi_2\}_{\hbox
{\scriptsize SD}},\qquad
dS_a=f^{-1}n_a d^{n-2}x, 
\end{equation}
where
\begin{equation}
\{ {\hat\xi}_1, {\hat\xi}_2 \}_{\hbox{\scriptsize SD}}^t ={\hat\xi}_1^a\partial_a{\hat\xi}_2^t - {\hat\xi}_2^a\partial_a{\hat\xi}_1^t,\qquad
\{ {\hat\xi}_1, {\hat\xi}_2 \}_{\hbox{\scriptsize SD}}^a =
{\hat\xi}_1^b\partial_b{\hat\xi}_2^a - {\hat\xi}_2^b\partial_b{\hat\xi}_1^a
+ g^{ab}\left({\hat\xi}_1^t\partial_b{\hat
\xi}_2^t-
{\hat\xi}_2^t\partial_b{\hat\xi}_1^t \right).
\end{equation}
Now in the above we can take into account the boundary conditions from paper \cite{Carlip}. Then we get, for example,
\begin{equation}
\frac{1}{8\pi G}\sqrt{g}(g^{ab}g^{cd}-g^{ac}g^{bd})\left(\hat\xi_{1a}\hat\xi^t_{2|bd}-\hat\xi_{
2a}\hat\xi^t_{1|bd}\right)f^{-1}n_c=\frac{1}{8\pi G}f\sqrt{\sigma}\sigma^{\alpha\beta}
\left(\hat\xi_2^r\hat\xi^t_{1|\alpha\beta}
- \hat\xi_1^r\hat\xi^t_{2|\alpha\beta}
\right)=O(N^2),
\end{equation}
here we suppose $\hat\xi^\alpha=0$ instead of  $\hat\xi^\alpha=O(1)$ because this condition is used in Section 3 of \cite{Carlip}.
In the similar way it occurs that all other integrands decrease as $O(N)$, or faster, i.e. they are zero on the black hole horizon. Therefore, the Regge-Teitelboim approach does not allow to derive expression (\ref{eq:2.13}) which is (2.13) of paper \cite{Carlip}. This is absolutely natural, because conditions for applying this method are just violated --- variation $\delta f$, induced by a hypersurface deformation, does not fulfil  Carlip's restriction   $\delta f/ f = O(N)$ and really has a form
\[
\delta_{\hat\xi} f=\left( f\hat\xi^r\right)_{,r}=O(N^{-1})\sim f,
\]
that is,   $\delta f/ f = O(1)$.
Due to this reason the surface integral from equation (2.10) of paper \cite{Carlip}
\begin{equation}
\delta L[{\hat\xi}] = \int_\Sigma d^{n-1}x\left(\frac{\delta H}{\delta
g_{ab}}\delta g_{ab}+\frac{\delta H}{\delta\pi^{ab}}\delta\pi^{ab} \right)
+ {1\over8\pi G} \oint_{r=r_+} d^{n-2}x  \delta n^r\partial_r{\hat\xi}^t
\sqrt{\sigma}.
\end{equation}
is nonzero, that should be qualified in Regge-Teitelboim's terminology as ``non-differentiability'' of functional $L[{\hat\xi}] =
H[{\hat\xi}]+J[{\hat\xi}]$, where
\begin{equation}
J[{\hat\xi}] = {1\over8\pi G}\oint_{r=r_+}\!\! d^2x\, \Bigl\{
n^a\nabla_a{\hat\xi}^t\sqrt{\sigma}
+ {\hat\xi}^a\pi_a{}^r \Bigr\}.
\end{equation}
Here we omit term $n_a{\hat\xi}^a K\sqrt{\sigma}$, which is present in
\cite{Carlip}. As it was told above we consider the inclusion of this term as a mistake which is incorrectly justified in \cite{Carlip} by supposing $\delta K_{rr}/K_{rr}=O(N)$.

\section{The New Poisson Brackets}
There is a more general  Hamiltonian approach where all functionals are admissible, including those giving nonzero boundary contribution to the first variation.
It was for the first time shown in publication \cite{LMMR} that the standard Poisson bracket can be generalized by adding to it some divergence terms.
As a result, the bracket so generalized can generate variations with nonzero surface contribution. This is just required in the case under consideration when the surface is a black hole horizon.
 Let us mention that two different extensions of the formula proposed in \cite{LMMR} were suggested later. They can be important in cases when  surface contributions to functional variations contain arbitrary  spatial derivatives for variations of
 fields and  their conjugate momenta (generally speaking,  arbitrary non-canonical variables): \cite{Sol93}, \cite{Bering}.
 An attempt of comparison of these two different formulas was done in paper \cite{antiBering}, but here all  give one and the same result.

 Let us explain this in more detail. Let field variables are not necessarily canonical, but their Poisson bracket is ultralocal
 \[
\{\phi_A(x),\phi_B(y)\}=I_{AB}\delta(x,y),
\]
then the standard bracket of two local functionals has the following form
\begin{equation}
\{F,G\}=\int_\Sigma d^{n-1}x\frac{\delta F}{\delta\phi_A(x)}I_{AB}\frac{\delta
G}{\delta\phi_B(x)},\label{eq:PB_old} 
\end{equation}
and for a ``differentiable'' functional we have \[
\delta F=\int_\Sigma d^{n-1}x\frac{\delta F}{\delta\phi_A(x)}\delta\phi_A(x).
\]
Let now  deal with ``non-differentiable'' functional in Regge-Teitelboim's sense, and let its first variation  has the simplest possible form with  a surface term
 \[
\delta F=\int_\Sigma d^{n-1}x\frac{\delta^\wedge
F}{\delta\phi_A(x)}\delta\phi_A(x)+ \oint_{\partial\Sigma}
d^{n-2}x\frac{\delta^\vee F}{\delta\phi_A(x)}\delta\phi_A(x),
\]
here we consider the above expression as a definition of $\delta^\wedge$ and $\delta^\vee$. These notatations are directly taken from \cite{LMMR}.
Then  we can formally construct a full variational derivative as follows
\[
\frac{\delta F}{\delta\phi_A(x)}=\theta(\Sigma)\frac{\delta^\wedge
F}{\delta\phi_A(x)}+\delta(\partial\Sigma)\frac{\delta^\vee
F}{\delta\phi_A(x)}, \]
to include a surface (or boundary) contribution (here
 $\delta(\partial\Sigma)$ is the Dirac delta-function concentrated on the boundary of the  integration region  $\Sigma$ and $\theta(\Sigma)$ is the Heaviside step function equal to $1$ inside $\Sigma$ and to $0$ outside of it). So, we treat the full variational derivative as follows
\[
\int_{{\bf R}^{n-1}} d^{n-1}x\frac{\delta
F}{\delta\phi_A(x)}\delta\phi_A(x)\equiv \int_\Sigma d^{n-1}x
\frac{\delta^\wedge F}{\delta\phi_A(x)}\delta\phi_A(x)+
\oint_{\partial\Sigma}d^{n-2}x \frac{\delta^\vee
F}{\delta\phi_A(x)}\delta\phi_A(x). \] It is possible to put these
full variational derivatives into the standard formula for the
Poisson bracket (\ref{eq:PB_old}) instead of the usual
Euler-Lagrange derivatives. Then we get
 \[
\{F,G\}_{\rm new}=\{F,G\}_{\rm old}+\oint_{\partial\Sigma}
d^{n-2}x\frac{\delta^\vee F}{\delta\phi_A}\delta_G\phi_A-
\oint_{\partial\Sigma}
d^{n-2}x\frac{\delta^\vee G}{\delta\phi_A}\delta_F\phi_A+?!, \]
where
\[
\delta_F\phi_A=I_{AB}\frac{\delta^\wedge F}{\delta\phi_B}=\{\phi_A,F\},\qquad
\delta_G\phi_A=I_{AB}\frac{\delta^\wedge G}{\delta\phi_B}=\{\phi_A,G\},
\]
and ?! denotes the puzzling term corresponding to the delta-function squared. In publication \cite{LMMR} it was demanded to kill this term by a boundary condition which guarantees zero  coefficient before the dangerous expression
$[\delta(\partial\Sigma)]^2$:
\[
\left(\frac{\delta^\vee F}{\delta\phi_A(x)}I_{AB}\frac{\delta^\vee
G}{\delta\phi_B(x)} \right)_{\partial\Sigma}=0. \]
Further attempts to extend the result of publication \cite{LMMR} were gone in two directions: 
\begin{enumerate}
 \item searching for regular expressions corresponding to terms like ?! \cite{Sol93};
\item postulating that these terms are absent in the final formula \cite{Bering}. \end{enumerate}

 Therefore, the contribution to expression (\ref{eq:2.13}) (which is (2.13) of paper \cite{Carlip}) arises just due to ``non-differentiability'' of functional $L[\hat\xi]$
\[
\delta L[\hat\xi]=\int_\Sigma d^{n-1}x\left(\frac{\delta H}{\delta
g_{ab}}\delta g_{ab}+\frac{\delta H}{\delta\pi^{ab}}\delta\pi^{ab} \right)
-\frac{1}{8\pi G}\oint_{r=r_+} d^2x  \sqrt{\sigma} \frac{\delta
f}{f^2}\hat\xi^t_{,r}\neq 0.
\]
The variation $\delta_{\hat\xi} f$ is easily determined from the equations of motion generated by the Hamiltonian  $L[\hat\xi]$
\begin{equation}
\delta_{\hat\xi}
f=\{f,L[\hat\xi]\}=\dot{\sqrt{g_{rr}}}
=\frac{\hat\xi_{r|r}}{f}=f\left(\hat\xi^r_
{,r}+\Gamma^r_{rr}\hat\xi^r \right)=\left( f\hat\xi^r\right)_{,r}.
\end{equation}
By this way we get the first surface contribution to the Poisson bracket  $\{L[\hat\xi_1],L[\hat\xi_2]\}_{\rm new}$ standing in expression  (\ref{eq:2.13})
\begin{equation}
-\frac{1}{8\pi G}\oint_{r=r_+} d^2x\frac{\sqrt{\sigma}}{f^2}\left[
\delta_{\hat\xi_2}f\hat\xi^t_{1,r}-\delta_{\hat\xi_1}f\hat\xi^t_{2,r}\right]=-\frac{1}{8\pi G}\oint_{r=r_+}
d^2x\frac{\sqrt{\sigma}}{f^2}\left[
\left(f\hat\xi^r_2\right)_{,r}\hat\xi^t_{1,r}-\left(f\hat\xi^r_1\right)_{,r}\hat
\xi^t_{2,r}\right]. \end{equation}
The genesis of the second term of expression (\ref{eq:2.13})
\begin{equation}
-{1\over8\pi G} \oint_{r=r_+} d^2x\frac{\sqrt{\sigma}}{f}\partial_r
\left[
\delta_{{\hat\xi}_1}{\hat\xi}_2^t
- \delta_{{\hat\xi}_2}{\hat\xi}_1^t \right]=
-{1\over8\pi G} \oint_{r=r_+}  d^2x\frac{\sqrt{\sigma}}{f}\partial_r
\left[
{\hat\xi}_1^r\partial_r{\hat\xi}_2^t - {\hat\xi}_2^r\partial_r{\hat\xi}_1^t
\right],
\end{equation}
is connected with the fact that the deformation parameters
$\hat\xi^t$ themselves are dependent on the canonical variables
and so themselves have nonzero Poisson brackets with the Hamiltonian. When these parameters come
as multipliers before constraints this fact leads simply to changes in
the final constraint multipliers and it is not important for us as we
are on the constraint surface. But when they come in the
nonzero surface integrals  their variations
$\delta_{{\hat\xi}_2}{\hat\xi}_1^t =
{\hat\xi}_2^a\partial_a{\hat\xi}_1^t$ will give a
contribution to the Poisson bracket.

\section{Conclusion}
We have shown that the validity of results of paper \cite{Carlip} is in fact based on the application of the new formula for Poisson brackets. This is unfortunately not clear from the argumentation given in  \cite{Carlip}.  Formulas (\ref{eq:7}) (which is (2.11) of  \cite{Carlip}) and (\ref{eq:8}) are not valid if the Poisson bracket is defined by the standard expression.  So, we have one more testimony in favor of the move
to transcend the Regge-Teitelboim approach in studying the role of boundaries in the Hamiltonian dynamics \cite{Sol93}, \cite{SolPR},
\cite{SolTMF}. 

Recently there appeared a new paper by Carlip \cite{Carlip2} where the difficulties discussed here are avoided by using the covariant canonical formalism. It maybe also of interest to study  the role of surface terms in Poisson brackets for this new approach.

The author is grateful to S.N.~Solodukhin for his e-mail comment given at the beginning of this work and to the referee for valuable critics.

\end{document}